\documentclass[10pt,twocolumn,amssymb,amsmath,aps,pra]{revtex4-1}
\usepackage{graphicx}
\usepackage{array}

\begin{document}

\title{Lorentz covariance of the mass-polariton theory of light}
\date{March 27, 2019}
\author{Mikko Partanen}
\author{Jukka Tulkki}
\affiliation{Engineered Nanosystems Group, School of Science, Aalto University, P.O. Box 12200, 00076 Aalto, Finland}

\begin{abstract}
In the mass-polariton (MP) theory of light formulated by us recently [Phys.~Rev.~A 95, 063850 (2017)], light in a medium is described as a coupled state of the field and matter. The key result of the MP theory is that the optical force of light propagating in a transparent material drives forward an atomic mass density wave (MDW). In previous theories, it has been well understood that the medium carries part of the momentum of light. The MP theory is fundamentally different since it shows that this momentum is associated with the MDW that carries a substantial atomic mass density and the related rest energy with light. In this work, we prove the Lorentz covariance of the MP theory and show how the stress-energy-momentum (SEM) tensor of the MP transforms between arbitrary inertial frames. We also compare the MP SEM tensor with the conventional Minkowski SEM tensor and show how the well-known fundamental problems of the Minkowski SEM tensor become solved by the SEM tensor based on the MP theory. We have particularly written our work for non-expert readers by pointing out how the Lorentz transformation and various conservation laws and symmetries of the special theory of relativity are fulfilled in the MP theory.
\end{abstract}

\maketitle

\section{Introduction}

We have recently introduced the mass-polariton (MP) theory of light \cite{Partanen2017c,Partanen2017e,Partanen2018a,Partanen2018b}, which differs from all previous theories of light by describing light as a coupled state of the field and an atomic mass density wave (MDW), which is driven forward by the optical force. Although many previous theories acknowledge the presence of the momentum of the medium \cite{Barnett2010b,Barnett2010a,Milonni2010,Hinds2009,Pfeifer2007,Brevik1979,Brevik2018b,Saldanha2017}, they all neglect the transfer of mass and the related rest energy by the MDW. Neglecting this transfer of mass and rest energy leads to an unavoidable contradiction with the conservation laws of nature and breaks the covariance principle of the special theory of relativity (STR). The shift of atoms with the MDW predicted by the MP theory of light is experimentally verifiable. It provides a complementary approach to discover the momentum of light in different media, and thus, may revive the experimental studies of the Abraham-Minkowski controversy \cite{Astrath2014,Ashkin1973,Casner2001,Pozar2018,Choi2017,Jones1954,Jones1978,Walker1975,She2008,Zhang2015,Campbell2005,Sapiro2009}.

In our previous work \cite{Partanen2017c}, the stress-energy-momentum (SEM) tensor formulation of the MP theory was discussed only briefly in Appendix B and mainly in the rest frame of the medium. In this work, we discuss in detail how the SEM tensor of the MP theory transforms between arbitrary inertial frames. Thus, the present work complements our resolution of the Abraham-Minkowski controversy
\cite{Partanen2017c,Leonhardt2006a,Cho2010,Kemp2017,Kemp2015,Bliokh2017a,Bliokh2017b,Penfield1967} by giving detailed space-time considerations of the MP theory within the framework of the STR. We also compare the SEM tensor of the MP theory with the conventional Minkowski SEM tensor. In particular, we will show how excluding the atomic MDW part from the SEM tensor leads to inconsistencies in fulfilling the covariance properties and the conservation laws that are built-in in the correctly formulated SEM tensor in the STR. Throughout this work, we make direct transparent reference to the fundamental definitions of concepts in the STR: Lorentz transformation, four-vector, SEM tensor, and the covariance principle as they are described, for instance, in the well-known textbook of Landau and Lifshitz in Ref.~\cite{Landau1989}.

This paper is organized as follows: Section~\ref{sec:tensors} describes the SEM tensor formulation of the MP theory of light. The relation of the SEM tensor of the MP theory to the conservation laws is discussed in Sec.~\ref{sec:conservation}. Section \ref{sec:covariance} describes the covariance properties of the MP theory, including the Lorentz transformation of the SEM tensor components and the covariant form of the field and the MDW equations. Section \ref{sec:comparison} presents the comparison of the MP theory of light with the conventional Minkowski SEM tensor formulation. The key results are represented in Table \ref{tbl:comparison}. Finally, conclusions are drawn in Sec.~\ref{sec:conclusions}. This paper is not aimed to be a balanced review of the SEM tensor formalisms of light, but it introduces the SEM tensor of the MP theory of light and proves its covariance properties. Comparison to other theoretical approaches is limited to the Minkowski SEM tensor.

\section{\label{sec:tensors}SEM tensor in the laboratory frame}

\subsection{Concepts and approximations}

For simplicity, in this work, we assume that the medium is nondispersive, linear, and isotropic. Even with these restrictions, the theory covers a broad area of optical phenomena and applications in photonics technologies in solids, liquids, and gases. Our concepts can also be extended in a slightly more complex form to more general media.

It is well known that any SEM tensor that conserves angular momentum must be symmetric \cite{Landau1989,Jackson1999,Misner1973}. The conventional general form of a symmetric SEM tensor in the Minkowski space-time is given by \cite{Landau1989}
\begin{equation}
 \mathbf{T}=
 \left[\begin{array}{cc}
  W & c\mathbf{G}^T\\
  c\mathbf{G} & \boldsymbol{\mathcal{T}}\\
 \end{array}\right]
 =\left[\begin{array}{cccc}
  W & cG^x & cG^y & cG^z\\
  cG^x & \mathcal{T}^{xx} & \mathcal{T}^{xy} & \mathcal{T}^{xz}\\
  cG^y & \mathcal{T}^{yx} & \mathcal{T}^{yy} & \mathcal{T}^{yz}\\
  cG^z & \mathcal{T}^{zx} & \mathcal{T}^{zy} & \mathcal{T}^{zz}
 \end{array}\right],
 \label{eq:emt}
\end{equation}
where $W$ is the energy density, $\mathbf{G}=(G^x,G^y,G^z)$ is the momentum density, and $\boldsymbol{\mathcal{T}}$ is the stress tensor with components $\mathcal{T}^{jk}$, where $j,k\in\{x,y,z\}$.

The SEM tensor of the MP and its electromagnetic field and the atomic MDW parts were originally presented in Appendix B of Ref.~\cite{Partanen2017c} in the laboratory frame (L frame), where the medium atoms are at rest (excluding possible thermal motion) before the optical force starts to accelerate them. We start with a brief introduction to these tensors in the L frame before investigating how these tensors transform between arbitrary inertial frames in Sec.~\ref{sec:covariance}.

\subsection{Elastic energy and relaxation}

In this work, we consider the SEM tensor of light in the infinite-medium limit where the vacuum-medium interfaces are not accounted for. We also assume that the dynamical variables of the medium do not appear in the field part of the SEM tensor. We know from the computer simulations based on the optoelastic continuum dynamics (OCD) \cite{Partanen2017c} that the loss of the field energy caused by the field-driven MDW dynamics is very small, although not exactly zero.

Thus, propagation of light in a medium is described with good accuracy, even if we neglect any strain energies that are left in the medium after a light wave. These strain energies are important in the description of the relaxation dynamics of the medium \cite{Partanen2017c,Partanen2017e,Partanen2018b}, but they are negligible in comparison with the field energy. It is obvious that the density of elastic energy that is generated in the medium by light could be added to the SEM tensor description below, but it is left as a topic for future work.

In the present work, we do not focus on the dynamical equation of the medium that has been described earlier for solid dielectrics \cite{Partanen2017c,Partanen2017e,Partanen2018a,Partanen2018b}. Instead, from the perspective of the medium dynamics, we only account for the atomic MDW effect, which is driven forward by the optical force density and is essentially independent of other terms in the dynamical equation of the medium.

Using these assumptions, the total SEM tensor of the MP is a sum of the SEM tensor $\mathbf{T}_\mathrm{field}$ of the electromagnetic field, which includes the energy related to the polarization of the material, and the SEM tensor $\mathbf{T}_\mathrm{MDW}$ of the field-driven atomic MDW as \cite{Partanen2017c}
\begin{equation}
 \mathbf{T}_\mathrm{MP}=\mathbf{T}_\mathrm{field}+\mathbf{T}_\mathrm{MDW}.
 \label{eq:tensorsum}
\end{equation}

\subsection{SEM tensor of the electromagnetic field}

In the SEM tensor of the electromagnetic field, we use the well-known electromagnetic energy density $W_\mathrm{field}$, the momentum density $\mathbf{G}_\mathrm{field}$, and the stress tensor $\boldsymbol{\mathcal{T}}_\mathrm{field}$, given in terms of the electric field $\mathbf{E}$, magnetic field $\mathbf{H}$, electric flux density $\mathbf{D}$, and magnetic flux density $\mathbf{B}$ by \cite{Jackson1999,Landau1984}
\begin{equation}
 W_\mathrm{field}=\frac{1}{2}(\mathbf{E}\cdot\mathbf{D}+\mathbf{H}\cdot\mathbf{B}),
 \label{eq:energyfield}
\end{equation}
\begin{equation}
 \mathbf{G}_\mathrm{field}=\frac{\mathbf{E}\times\mathbf{H}}{c^2},
 \label{eq:momentumfield}
\end{equation}
\begin{equation}
 \boldsymbol{\mathcal{T}}_\mathrm{field}=\frac{1}{2}(\mathbf{E}\cdot\mathbf{D}+\mathbf{H}\cdot\mathbf{B})\mathbf{I}-\mathbf{E}\otimes\mathbf{D}-\mathbf{H}\otimes\mathbf{B}.
\label{eq:electromagneticstress}
\end{equation}
Here, $\otimes$ denotes the outer product and $\mathbf{I}$ is the $3\times3$ unit matrix. Note that $\boldsymbol{\mathcal{T}}_\mathrm{field}$ is \emph{generally asymmetric} in an arbitrary inertial frame. However, this is not a problem since the total SEM tensor of the MP, which is a sum of the field and the MDW parts in Eq.~\eqref{eq:tensorsum}, will be \emph{symmetric in all inertial frames} as described in Sec.~\ref{sec:covariance}.

Note that, in the present work, we use the conventional constitutive relations $\mathbf{D}=\varepsilon\mathbf{E}$ and $\mathbf{B}=\mu\mathbf{H}$ in the L frame, where $\varepsilon$ and $\mu$ are the permittivity and permeability of the medium. In any other inertial frames, the relations between the field quantities are more complicated, but they are unambiguously tied to the relations in the L frame by the well-known Lorentz transformations of the fields described in Sec.~\ref{sec:covarianceeq}.

By substituting the energy density, momentum density, and the stress tensor from Eqs.~\eqref{eq:energyfield}--\eqref{eq:electromagneticstress}
into the general form of a SEM tensor in Eq.~\eqref{eq:emt}, we obtain the SEM tensor of the electromagnetic field, given by
\begin{align}
 &\mathbf{T}_\mathrm{field}\nonumber\\
 &\!\!=\!
 \bigg[\begin{array}{cc}
  \!\!\frac{1}{2}(\mathbf{E}\!\cdot\!\mathbf{D}\!+\!\mathbf{H}\!\cdot\!\mathbf{B}) & \frac{1}{c}(\mathbf{E}\!\times\!\mathbf{H})^T\\
  \frac{1}{c}\mathbf{E}\!\times\!\mathbf{H} & \frac{1}{2}(\mathbf{E}\!\cdot\!\mathbf{D}\!+\!\mathbf{H}\!\cdot\!\mathbf{B})\mathbf{I}-\mathbf{E}\!\otimes\!\mathbf{D}-\mathbf{H}\!\otimes\!\mathbf{B}\!\!
 \end{array}\bigg].
 \label{eq:tensorfield}
\end{align}
This is conventionally known as the Abraham SEM tensor \cite{Pfeifer2007}. We also note that the trace of the energy momentum tensor of the field in Eq.~\eqref{eq:tensorfield} is zero, which is related to the masslessness of the electromagnetic field \cite{Jackson1999}.

\subsection{SEM tensor of the atomic MDW}

In the MP theory of light, we apply the SEM tensor given in Eq.~\eqref{eq:tensorfield} for the electromagnetic field. Due to the Abraham force density $\mathbf{f}_\mathrm{A}=\frac{\partial}{\partial t}(\mathbf{D}\times\mathbf{B}-\mathbf{E}\times\mathbf{H}/c^2)$, light propagating in a medium drives forward an atomic MDW, which essentially disturbs the SEM tensor of the matter from its equilibrium value. The SEM tensor $\mathbf{T}_\mathrm{MDW}$ of the atomic MDW is obtained as the difference of the actual SEM tensor $\mathbf{T}_\mathrm{mat,a}$ of the matter and the SEM tensor $\mathbf{T}_\mathrm{mat,0}$ of the matter in the absence of light. 

Using the well-known expression of the SEM tensor of the mass density of the matter \cite{Misner1973,Landau1987}, the SEM tensor of the MDW is given in the L frame by \cite{Partanen2017c}
\begin{align}
\mathbf{T}_\mathrm{MDW} &=\mathbf{T}_\mathrm{mat,a}-\mathbf{T}_\mathrm{mat,0}\nonumber\\
 &=\left[\begin{array}{cccc}
\rho_\mathrm{a}c^2 & \rho_\mathrm{a}\mathbf{v}_\mathrm{a}^Tc\\
\rho_\mathrm{a}\mathbf{v}_\mathrm{a}c & \rho_\mathrm{a}\mathbf{v}_\mathrm{a}\otimes\mathbf{v}_\mathrm{a}
\end{array}\right]
-\left[\begin{array}{cccc}
\rho_0c^2 & \mathbf{0}\\
\mathbf{0} & \mathbf{0}
\end{array}\right]\nonumber\\
 &=\bigg[\begin{array}{cc}
  \rho_\mathrm{MDW}c^2 & \rho_\mathrm{MDW}\mathbf{v}_\mathrm{l}^Tc\\
  \rho_\mathrm{MDW}\mathbf{v}_\mathrm{l}c & \rho_\mathrm{MDW}\mathbf{v}_\mathrm{a}\otimes\mathbf{v}_\mathrm{l}
 \end{array}\bigg].
 \label{eq:tensormdw}
\end{align}
where $\mathbf{v}_\mathrm{a}$ is the local atomic velocity in the MDW, $\mathbf{v}_\mathrm{l}$ is the local velocity of light in the medium, $\rho_0$ is the atomic mass density in the absence of the MDW, and $\rho_\mathrm{a}$ is the actual mass density of atoms. Note that $\rho_\mathrm{a}$ differs from $\rho_0$ due to the density variations of atoms caused by the Abraham force. The excess mass density of atoms in the MDW is then given by $\rho_\mathrm{MDW}=\rho_\mathrm{a}-\rho_0$ and it also satisfies $\rho_\mathrm{MDW}\mathbf{v}_\mathrm{l}=\rho_\mathrm{a}\mathbf{v}_\mathrm{a}$, which one can show analytically in the case of an electromagnetic plane wave and by computer simulations for a general light pulse \cite{Partanen2017c}.

From the relation $\rho_\mathrm{MDW}\mathbf{v}_\mathrm{l}=\rho_\mathrm{a}\mathbf{v}_\mathrm{a}$, one can see that the momentum density of the MDW is the classical momentum density of the medium. From Eq.~\eqref{eq:tensormdw}, we then obtain the energy and momentum densities and the stress tensor of the MDW, given by \cite{Partanen2017c}
\begin{equation}
 W_\mathrm{MDW}=\rho_\mathrm{a}c^2-\rho_0c^2=\rho_\mathrm{MDW}c^2,
 \label{eq:energymdw}
\end{equation}
\begin{equation}
 \mathbf{G}_\mathrm{MDW}=\rho_\mathrm{a}\mathbf{v}_\mathrm{a}=\rho_\mathrm{MDW}\mathbf{v}_\mathrm{l},
 \label{eq:momentummdw}
\end{equation}
\begin{equation}
 \boldsymbol{\mathcal{T}}_\mathrm{MDW}=\rho_\mathrm{a}\mathbf{v}_\mathrm{a}\otimes\mathbf{v}_\mathrm{a}=\rho_\mathrm{MDW}\mathbf{v}_\mathrm{a}\otimes\mathbf{v}_\mathrm{l}.
 \label{eq:stressmdw}
\end{equation}
In a moving reference frame, one must subtract from the actual atomic momentum, the momentum of $\rho_0$, which is no longer at rest.

Note that the last form of Eq.~\eqref{eq:tensormdw} and the right-hand sides of Eqs.~\eqref{eq:energymdw}--\eqref{eq:stressmdw} are the general expressions of the MDW quantities, which extend to arbitrary inertial frames as described in Sec.~\ref{sec:covariance}. In the special case of the L frame, where the velocity of the MDW atoms is negligible, the rest energy density of the MDW can be expressed in terms of the field quantities as \cite{Partanen2017c}
\begin{equation}
 W_\mathrm{MDW}^{(\mathrm{L})}\approx\frac{n^2-1}{2}(\mathbf{E}\cdot\mathbf{D}+\mathbf{H}\cdot\mathbf{B}),
\label{eq:mdwrestframeenergy}
\end{equation}
where $n$ is the refractive index. The MDW momentum density in the L frame can be written as \cite{Partanen2017c}
\begin{equation}
\mathbf{G}_\mathrm{MDW}^{(\mathrm{L})}\approx\mathbf{D}\times\mathbf{B}-\frac{\mathbf{E}\times\mathbf{H}}{c^2}=\frac{n^2-1}{c^2}\mathbf{E}\times\mathbf{H}.
\label{eq:mdwrestframemomentum}
\end{equation}
The kinetic energy terms in the MDW stress tensor in Eq.~\eqref{eq:stressmdw} are negligibly small in the L frame due to the second-order dependence on the small atomic velocity $\mathbf{v}_\mathrm{a}$ \cite{Partanen2017c}. Thus, we have
\begin{equation}
\boldsymbol{\mathcal{T}}_\mathrm{MDW}^{(\mathrm{L})}\approx\mathbf{0}.
\label{eq:mdwrestframestress}
\end{equation}However, the MDW stress tensor generally becomes essential in moving reference frames as described in Sec.~\ref{sec:covariance}. It is found to maintain the total stress tensor of the field and the MDW symmetric in all inertial frames. Note that Eqs.~\eqref{eq:mdwrestframeenergy}--\eqref{eq:mdwrestframestress} have been made possible by the assumption that the back-action of the field-driven medium dynamics on the field is negligible \cite{Partanen2017c}.

\subsection{Mass-energy equivalence in the STR}

It is obvious that the absence of the rest energy of the atoms in the MDW moving with light is a fundamental problem in earlier SEM tensor formulations of light. Therefore, we briefly summarize how the total energy of particles is treated in the STR. The general expression for the total energy of a particle is given by $\gamma m_0c^2$, where $m_0$ is the rest mass and $\gamma=1/\sqrt{1-v^2/c^2}$ is the Lorentz factor. The particle energy can be split into the rest energy equal to $m_0c^2$ and kinetic energy equal to $(\gamma-1) m_0c^2$. The particle velocity, the velocity of the medium atoms in our case, is $v_\mathrm{a}\ll c$, and the kinetic energy can be written in the nonrelativistic form $\frac{1}{2}m_0v_\mathrm{a}^2$. In the L frame, we correspondingly have the rest energy density $\rho_\mathrm{a}c^2$, the kinetic energy density $\frac{1}{2}\rho_\mathrm{a}v_\mathrm{a}^2$, and the corresponding energy fluxes. Thus, atoms moving in a medium driven by the optical force of the field carry both their kinetic energy and rest energy, and the corresponding energy fluxes must be added to the energy flux of the field to obtain the total energy flux. The same energy flux consideration also proves that the rest energy density of the MDW must be included in the total SEM tensor of light.

\subsection{\label{sec:tensormp}SEM tensor of the coupled MP state}

By substituting the SEM tensor of the electromagnetic field in Eq.~\eqref{eq:tensorfield} and the SEM tensor of the atomic MDW in Eq.~\eqref{eq:tensormdw} into Eq.~\eqref{eq:tensorsum}, we then obtain the total SEM tensor of the MP as
\begin{align}
 &\mathbf{T}_\mathrm{MP}\nonumber\\
 &\!=\!\!\bigg[\begin{array}{cc}
  \!\!\frac{1}{2}(\mathbf{E}\!\cdot\!\mathbf{D}+\mathbf{H}\!\cdot\!\mathbf{B})+\rho_\mathrm{MDW}c^2 & \frac{1}{c}(\mathbf{E}\!\times\!\mathbf{H})^T\!+\rho_\mathrm{MDW}\mathbf{v}_\mathrm{l}^Tc\!\\
  \frac{1}{c}\mathbf{E}\!\times\!\mathbf{H}+\rho_\mathrm{MDW}\mathbf{v}_\mathrm{l}c & \boldsymbol{\mathcal{T}}_\mathrm{MP}
 \end{array}\bigg],
\label{eq:tensormp}
\end{align}
where the MP stress tensor $\boldsymbol{\mathcal{T}}_\mathrm{MP}=\boldsymbol{\mathcal{T}}_\mathrm{field}+\boldsymbol{\mathcal{T}}_\mathrm{MDW}$ is given by
\begin{equation}
 \boldsymbol{\mathcal{T}}_\mathrm{MP}=\frac{1}{2}(\mathbf{E}\cdot\mathbf{D}+\mathbf{H}\cdot\mathbf{B})\mathbf{I}-\mathbf{E}\otimes\mathbf{D}-\mathbf{H}\otimes\mathbf{B}+\rho_\mathrm{MDW}\mathbf{v}_\mathrm{a}\otimes\mathbf{v}_\mathrm{l}.
\label{eq:stress}
\end{equation}
The MP SEM tensor in Eq.~\eqref{eq:tensormp} is the total SEM tensor of light in the MP theory. In Sec.~\ref{sec:covariance}, we will show that this expression of the MP SEM tensor is form-invariant and it transforms according to the Lorentz transformation for second-rank tensors between arbitrary inertial frames.

\subsection{Angular momentum tensor}

The SEM tensor can also be used to describe angular momentum. In this section, we will review for completeness the description of the angular momentum density (AMD) tensor and the related angular momentum (AM) tensor. Using the index notation, where the indices $\alpha$, $\beta$, and $\gamma$ range over all four components $(ct,x,y,z)$ of the Minkowski space-time, the angular momentum density with respect to the origin is given by the third-rank AMD tensor \cite{Jackson1999,Landau1989,Misner1973}
\begin{equation}
 \mathcal{M}^{\alpha\beta\gamma}=x^\alpha T^{\beta\gamma}-x^\beta T^{\alpha\gamma},
 \label{eq:angularmomentumdensity}
\end{equation}
which is antisymmetric with respect to the indices $\alpha$ and $\beta$. The integral of the AMD tensor in Eq.~\eqref{eq:angularmomentumdensity} over the boundary $\partial\Omega$ of a four-dimensional space-time region $\Omega$ (i.e., $\partial\Omega$ is a three-dimensional space-time hypersurface) gives the second-rank AM tensor as \cite{Misner1973}
\begin{equation}
 M^{\alpha\beta}=\frac{1}{c}\oint_{\partial\Omega}\mathcal{M}^{\alpha\beta\gamma}d\Sigma_\gamma.
\end{equation}
Here the differential volume element $d\Sigma_\gamma$ is proportional to a four-dimensional unit vector that is normal to the three-dimensional space-time hypersurface. The integral is taken over the coordinates $x$. By choosing the hypersurface to be a spacelike surface of constant time, i.e., $\gamma=0$ and $d\Sigma_0=dxdydz=d^3r$, and assuming that the AMD tensor of an isolated system becomes zero at infinity, we obtain the total AM tensor of the system as \cite{Misner1973}
\begin{equation}
 M^{\alpha\beta}=\frac{1}{c}\int\mathcal{M}^{\alpha\beta 0}d^3r.
 \label{eq:angularmomentumtensor1}
\end{equation}
We denote $x=(ct,\mathbf{r})$, where $\mathbf{r}$ is a three-dimensional vector, and define in the conventional way the three-dimensional angular momentum density $\boldsymbol{\mathcal{J}}$ as \cite{Andrews2013,Piccirillo2013,Allen1992b,Jackson1999,Landau1984,Landau1989}
\begin{equation}
 \boldsymbol{\mathcal{J}}=\mathbf{r}\times\mathbf{G}.
\end{equation}
It is also convenient to define the related quantity $\boldsymbol{\mathcal{N}}$, which is, in the recent optics literature \cite{Bliokh2013b,Bliokh2018a,Smirnova2018}, called boost momentum by
\begin{equation}
 \boldsymbol{\mathcal{N}} =\frac{W}{c^2}\mathbf{r}+\mathbf{G}t.
\end{equation}
Conservation of the boost momentum ensures the rectilinear motion of the energy centroid of light in a homogeneous medium. In terms of the quantities $\boldsymbol{\mathcal{J}}$ and $\boldsymbol{\mathcal{N}}$, the AM tensor in Eq.~\eqref{eq:angularmomentumtensor1} can be expressed as a matrix \cite{Fayngold2008}
\begin{align}
 \mathbf{M} &=
 \int\left[\begin{array}{cc}
  0 & -c\boldsymbol{\mathcal{N}}^T\\
  c\boldsymbol{\mathcal{N}} & \mathbf{r}\wedge\mathbf{G}\\
 \end{array}\right]d^3r\nonumber\\
 &=\int\left[\begin{array}{cccc}
  0 & -c\mathcal{N}^x & -c\mathcal{N}^y & -c\mathcal{N}^z\\
  c\mathcal{N}^x & 0 & \mathcal{J}^z & -\mathcal{J}^y\\
  c\mathcal{N}^y & -\mathcal{J}^z & 0 & \mathcal{J}^x\\
  c\mathcal{N}^z & \mathcal{J}^y & -\mathcal{J}^x & 0
 \end{array}\right]d^3r,
 \label{eq:angularmomentumtensor2}
\end{align}
where $\wedge$ denotes the exterior product. Like the SEM tensor, the consistent AM tensor of an isolated system must be form-invariant and transform according to the Lorentz transformation for second-rank tensors as described in Sec.~\ref{sec:covariance}.

\section{\label{sec:conservation}Conservation laws and the SEM tensor}

\subsection{Conservation laws and the continuity equation of the atomic MDW}

Since the coupled state of the field and the MDW is an isolated system, its four-momentum  is conserved. The conservation law of four-momentum is well known and given, e.g., in Refs.~\cite{Penfield1967,Jackson1999,Landau1984}. In the case of a diagonally symmetric SEM tensor, this conservation law must be written as
\begin{equation}
 \frac{1}{c^2}\frac{\partial W}{\partial t}+\nabla\cdot\mathbf{G}=-\frac{\phi}{c^2},
 \label{eq:conservationphi}
\end{equation}
\begin{equation}
 \frac{\partial\mathbf{G}}{\partial t}+\nabla\cdot\boldsymbol{\mathcal{T}}=-\mathbf{f},
 \label{eq:conservationf}
\end{equation}
where $\mathbf{f}$ is the force density and $\phi$ is the power-conversion density, both of which can be set to zero for an isolated system. Then, Eq.~\eqref{eq:conservationphi} describes the conservation of energy and Eq.~\eqref{eq:conservationf} describes the conservation of momentum.

It is also important to note that the atomic MDW obeys the continuity equation
\begin{equation}
 \frac{1}{c^2}\frac{\partial}{\partial t}(\rho_\mathrm{MDW}c^2)+\nabla\cdot(\rho_\mathrm{MDW}\mathbf{v}_\mathrm{l})=0.
 \label{eq:continuity}
\end{equation}
Therefore, the MDW terms can be subtracted from the left-hand side of Eq.~\eqref{eq:conservationphi} without changing the right-hand side of this equation. In previous theoretical works, the MDW terms have not been included in Eq.~\eqref{eq:conservationphi} \cite{Kemp2017,Penfield1967}. This corresponds to writing this equation with substitutions $\mathbf{G}\rightarrow\mathbf{G}-\rho_\mathrm{MDW}\mathbf{v}_\mathrm{l}=\mathbf{G}_\mathrm{field}$ and $W\rightarrow W-\rho_\mathrm{MDW}c^2=W_\mathrm{field}$ as $\frac{1}{c^2}\frac{\partial}{\partial t}W_\mathrm{field}+\nabla\cdot\mathbf{G}_\mathrm{field}=-\frac{\phi}{c^2}$ \cite{Penfield1967,Jackson1999,Landau1984}. 

Even if the MDW terms could be neglected from Eq.~\eqref{eq:conservationphi}, these terms are of fundamental importance for the consistency of the total energy momentum tensor; see Sec.~\ref{sec:comparison}. The MDW terms also play an important role in the form invariance of the SEM tensor in the Lorentz transformation as will be described in Sec.~\ref{sec:covariance}.

Using the index notation and the Einstein summation convention, the conservation law of angular momentum can be written in terms of the AMD tensor in Eq.~\eqref{eq:angularmomentumdensity} as
\begin{equation}
 \partial_\gamma\mathcal{M}^{\alpha\beta\gamma}=0.
 \label{eq:angularmomentumconservation}
\end{equation}
It can be shown that this is linked to the diagonal symmetry of the SEM tensor \cite{Jackson1999,Landau1989,Misner1973}. Thus, all the conservation laws of energy, momentum, and angular momentum can be compactly written in terms of the SEM tensor as \cite{Jackson1999}
\begin{equation}
 \partial_\alpha T^{\alpha\beta}=0,\hspace{0.5cm} T^{\alpha\beta}=T^{\beta\alpha}.
 \label{eq:conservation}
\end{equation}
The first equation here corresponds to Eqs.~\eqref{eq:conservationphi} and \eqref{eq:conservationf} for an isolated system like the coupled MP state of the field and the MDW. For an isolated system, the four-force is zero. The second equation describes the diagonal symmetry needed to fulfill the conservation law of angular momentum in Eq.~\eqref{eq:angularmomentumconservation}. This equation must also be fulfilled for an isolated system.

The SEM tensor of the MP in Eq.~\eqref{eq:tensormp} obeys the conservation laws of energy, momentum, and angular momentum. Verifying that these conservation laws in Eq.~\eqref{eq:conservation} are satisfied for a field propagating in a medium with constant refractive index $n$ is straightforward using the MP SEM tensor in Eq.~\eqref{eq:tensormp} together with Eqs.~\eqref{eq:mdwrestframeenergy}--\eqref{eq:mdwrestframestress}, which apply in the L frame. The fulfillment of the conservation laws is also evident from the relations of Sec.~\ref{sec:Abraham} below. The conservation laws become automatically fulfilled for the MP SEM tensor in any other inertial frame since the MP SEM tensor transforms according to the Lorentz transformation as described in Sec.~\ref{sec:covariance}. This is a strong argument for the consistency of the MP theory of light since none of the other commonly used SEM tensors of light reproduce the conservation laws in all inertial frames of the space-time without introducing any artificial concepts such as artificial curved metrics \cite{Gordon1923,Leonhardt2006b,Leonhardt2010}.

\subsection{Notes on the four-force in lossy media}

For the case of a lossy medium, we make two notes on Eqs.~\eqref{eq:conservationphi} and \eqref{eq:conservationf}: (1) In the general non-isolated case, the four-force $(\phi/c,\mathbf{f})$ is the force experienced by the coupled MP state of the field and the atomic MDW. Therefore, $\mathbf{f}$ is not the force experienced by atoms, but the actual force on atoms is given by $\mathbf{f}_\mathrm{atoms}=\mathbf{f}+\frac{d}{dt}\mathbf{G}_\mathrm{MDW}$, where the second term describes the rate of change of the MDW momentum density and it is equal to the Abraham force described in Sec.~\ref{sec:Abraham} below. (2) In a lossy medium, the power-conversion density $\phi$ is related to the conversion of electromagnetic energy, since the atomic mass energy of the MDW coming from the rest energies of particles cannot be reduced and the kinetic energy of atoms was already assumed to be negligible. We do not consider the lossy medium case further.

\subsection{\label{sec:Abraham}Abraham force}

It is important to note that neither the field nor the MDW part of the total MP SEM tensor satisfies the conservation laws in Eq.~\eqref{eq:conservation}. This follows from the fact that the field and the atomic MDW parts of the total SEM tensor of the MP state are coupled by the Abraham force density
\begin{align}
 \mathbf{f}_\mathrm{A} &=-\frac{\partial\mathbf{G}_\mathrm{field}}{\partial t}-\nabla\cdot\boldsymbol{\mathcal{T}}_\mathrm{field}
 =\frac{\partial}{\partial t}\Big(\mathbf{D}\times\mathbf{B}-\frac{\mathbf{E}\times\mathbf{H}}{c^2}\Big)\nonumber\\
 &=\frac{\partial\mathbf{G}_\mathrm{MDW}}{\partial t}+\nabla\cdot\boldsymbol{\mathcal{T}}_\mathrm{MDW}=\frac{\partial}{\partial t}(\rho_\mathrm{MDW}\mathbf{v}_\mathrm{l}-\rho_\mathrm{MDW}\mathbf{v}_\mathrm{a})\nonumber\\
 &=\frac{d}{dt}(\rho_\mathrm{MDW}\mathbf{v}_\mathrm{l}).
 \label{eq:Abrahamforce2}
\end{align}
The first line gives the Abraham force in terms of the field quantities, and the second and third lines, give the Abraham force in terms of the MDW quantities.

\subsection{\label{sec:balance}Law of action and counteraction between the field and the MDW}

Since the coupled state of the field and the MDW is an isolated system, the external forces are absent, and the field and the matter parts of the total MP SEM tensor satisfy the dynamical equations of motion, given by
\begin{equation}
 \partial_\beta(T_\mathrm{field})^{\alpha\beta}=-(f_\mathrm{A})^\alpha,
 \label{eq:conservationfield}
\end{equation}
\begin{equation}
 \partial_\beta(T_\mathrm{MDW})^{\alpha\beta}=(f_\mathrm{A})^\alpha.
 \label{eq:conservationmdw}
\end{equation}
Thus, it immediately follows that the four-divergence of the total SEM tensor of the MP theory is zero since the Abraham force terms in Eqs.~\eqref{eq:conservationfield} and \eqref{eq:conservationmdw} cancel each other due to their opposite signs in the field and the MDW parts.

To summarize the properties of the SEM tensor presentation of the MP theory of light, we conclude that this tensor gives the dynamical equations of motion and all conservation laws in a consistent and transparent way. This makes it superior to all previously presented SEM tensor formalisms of light, which break in fulfilling all these key physical properties of a consistent physical theory. In the next section, we will show that the MP SEM tensor is form-invariant in a Lorentz transformation to an arbitrary inertial frame when the field and medium variables are transformed according to the Lorentz transformation. This will prove the Lorentz covariance of the MP theory.

\section{\label{sec:covariance}Lorentz covariance of the MP theory}

It is the fundamental requirement of the theory of relativity that any SEM tensor must be Lorentz-covariant. This requirement has two meanings, which are intimately linked to each other: (1) The components of the SEM tensors in different inertial frames must be unambiguously related to each other by the Lorentz transformation. (2) The SEM tensor must be written in terms of the Lorentz-covariant quantities, which hold the same form in all inertial frames. In other words, this means that the laws of physics must be the same for all inertial observers.

\subsection{\label{sec:Lorentzsecondrank}Lorentz transformation of the SEM tensor}

We assume that an arbitrary general inertial frame (G$'$ frame) is moving with respect to another arbitrary general inertial frame (G frame) with a constant velocity $\mathbf{v}$. In this general case, the Lorentz boost can be written in the matrix form as
\begin{equation}
 \boldsymbol{\Lambda}=\left[\begin{array}{cc}
  \gamma & -\gamma\frac{v}{c}\mathbf{n}^T\\
  -\gamma\frac{v}{c}\mathbf{n} & \mathbf{I}+(\gamma-1)\mathbf{n}\otimes\mathbf{n}
 \end{array}\right],
 \label{eq:Lorentzboost}
\end{equation}
where $v=|\mathbf{v}|$ is the magnitude of $\mathbf{v}$, $\gamma=1/\sqrt{1-v^2/c^2}$ is the Lorentz factor, and $\mathbf{n}=\mathbf{v}/v$ is the unit vector parallel to $\mathbf{v}$.

A second-rank tensor $\mathbf{T}$ in space-time transforms according to the Lorentz transformation as
\begin{equation}
 \mathbf{T}'=\boldsymbol{\Lambda}\mathbf{T}\boldsymbol{\Lambda}.
\label{eq:LorentzT}
\end{equation}
This condition unambiguously relates the tensor components in the G$'$ frame to those in the G frame. However, this condition alone does not make the SEM tensor Lorentz-covariant as the tensor components must also be written in terms of the Lorentz-covariant quantities to ensure the invariant form of the laws of physics in all inertial frames. This requirement in the case of the MP theory of light will be described in detail in Sec.~\ref{sec:covarianceeq}.

\subsection{\label{sec:covarianceeq}Lorentz covariance of the field and the MDW equations}

Next, we show that the total SEM tensor of the coupled MP state of the field and the MDW transforms in a Lorentz-covariant way from the G frame to the G$'$ frame. We utilize the Lorentz transformation of the electric and magnetic fields of the Minkowski form, given by \cite{Kemp2017,Penfield1967}
\begin{align}
 \mathbf{E}'&=\mathbf{E}_\parallel+\gamma(\mathbf{E}_\perp+\mathbf{v}\times\mathbf{B}),\label{eq:LorentzE}\\
 \mathbf{H}'&=\mathbf{H}_\parallel+\gamma(\mathbf{H}_\perp-\mathbf{v}\times\mathbf{D}),\label{eq:LorentzH}\\
 \mathbf{D}'&=\mathbf{D}_\parallel+\gamma(\mathbf{D}_\perp+\frac{1}{c^2}\mathbf{v}\times\mathbf{H}),\label{eq:LorentzD}\\
 \mathbf{B}'&=\mathbf{B}_\parallel+\gamma(\mathbf{B}_\perp-\frac{1}{c^2}\mathbf{v}\times\mathbf{E}).\label{eq:LorentzB}
\end{align}
The subscripts $\parallel$ and $\perp$ denote parallel and perpendicular components to the velocity $\mathbf{v}$. In addition, the MDW mass density, the velocity of light in the medium, and the atomic velocity in the MDW transform from the G frame to the G$'$ frame as
\begin{align}
 \rho_\mathrm{MDW}' &=\frac{c^2-\mathbf{v}_\mathrm{l}\cdot\mathbf{v}}{c^2-(\mathbf{v}\ominus\mathbf{v}_\mathrm{a})\cdot\mathbf{v}}\rho_\mathrm{MDW},\label{eq:Lorentzrho}\\
 \mathbf{v}_\mathrm{l}' &=-(\mathbf{v}\ominus\mathbf{v}_\mathrm{l}),\label{eq:Lorentzvl}\\
 \mathbf{v}_\mathrm{a}' &=-(\mathbf{v}\ominus\mathbf{v}_\mathrm{a}).\label{eq:Lorentzva}
\end{align}
where $\ominus$ denotes the relativistic velocity subtraction, defined for $\mathbf{v}$ and an arbitrary velocity vector $\mathbf{u}$ by the conventional relation \cite{Jackson1999}
\begin{equation}
 \mathbf{v}\ominus\mathbf{u}=\frac{1}{1-\frac{\mathbf{v}\cdot\mathbf{u}}{c^2}}\Big(\mathbf{v}-\frac{\mathbf{u}_\mathrm{\perp}}{\gamma}-\mathbf{u}_\mathrm{\parallel}\Big).
\end{equation}
Equations \eqref{eq:Lorentzrho}--\eqref{eq:Lorentzva} essentially separate the MP formulation of electrodynamics from the conventional Minkowski SEM theory. The MP formulation is also different from any other known formulation of electrodynamics, none of which presents the atomic MDW as an integral part of the total coupled state of light in a medium.

It is a straightforward technical task to check that the application of the field and the MDW transformation laws in Eqs.~\eqref{eq:LorentzE}--\eqref{eq:Lorentzva} to the MP SEM tensor in Eq.~\eqref{eq:tensormp} leads to the same MP SEM tensor in the G$'$ frame as the direct application of the Lorentz transformation in Eq.~\eqref{eq:LorentzT} to the MP SEM tensor in the G frame. The relative magnitudes of the field and the MDW contributions change between different inertial frames. To compare the magnitudes of the field and the MDW quantities, one must note that in the L frame, the MDW energy and momentum densities and the stress tensor can be expressed in terms of the field quantities as given in Eqs.~\eqref{eq:mdwrestframeenergy}--\eqref{eq:mdwrestframestress}. Therefore, the transformation laws in Eqs.~\eqref{eq:LorentzE}--\eqref{eq:Lorentzva} allow comparing the field and the MDW quantities also in an arbitrary inertial frame.

Correspondingly, one can also verify that the application of the transformations of the field and the MDW quantities in Eqs.~\eqref{eq:LorentzE}--\eqref{eq:Lorentzva} to the MP AM tensor defined through Eq.~\eqref{eq:angularmomentumtensor2} leads to the same MP AM tensor in the G$'$ frame as the direct application of the Lorentz transformation in Eq.~\eqref{eq:LorentzT} to the MP AM tensor in the G frame. In the latter case, if the Lorentz transformation in Eq.~\eqref{eq:LorentzT} is applied to the matrix in the integrand of Eq.~\eqref{eq:angularmomentumtensor2}, one must additionally transform the differential volume element.

We also note that neither the SEM tensor of the electromagnetic field in Eq.~\eqref{eq:tensorfield} nor the SEM tensor of the MDW in Eq.~\eqref{eq:tensormdw} alone satisfies the two requirements of the Lorentz covariance simultaneously. However, their sum, which is the total MP SEM tensor, satisfies the two requirements. This is an additional strong argument for considering the field and the MDW tensors as inseparable parts of the complete MP SEM tensor. The same discussion applies to the MP AM tensor.

\subsection{\label{sec:covarianceeqtot}Lorentz transformation of the total energy, momentum, and angular momentum of light}

In summary, the Lorentz-covariant expressions for the total energy, momentum, angular momentum, and boost momentum densities of light are given by
\begin{align}
 W_\mathrm{MP} &=\frac{1}{2}(\mathbf{E}\cdot\mathbf{D}+\mathbf{H}\cdot\mathbf{B})+\rho_\mathrm{MDW}c^2,\label{eq:Wmp}\\
 \mathbf{G}_\mathrm{MP} &=\frac{\mathbf{E}\times\mathbf{H}}{c^2}+\rho_\mathrm{MDW}\mathbf{v}_\mathrm{l},\label{eq:Gmp}\\
 \boldsymbol{\mathcal{J}}_\mathrm{MP} &=\mathbf{r}\times\mathbf{G}_\mathrm{MP},\label{eq:Jmp}\\
 \boldsymbol{\mathcal{N}}_\mathrm{MP} &=\frac{W_\mathrm{MP}}{c^2}\mathbf{r}+\mathbf{G}_\mathrm{MP}t\label{eq:Nmp}.
\end{align}
The first terms of Eqs.~\eqref{eq:Wmp} and \eqref{eq:Gmp} are the contributions of the electromagnetic field and the second terms are the contributions of the atomic MDW. Due to the linearity of Eqs.~\eqref{eq:Jmp} and \eqref{eq:Nmp}, also the angular and boost momenta can be split into the field and the MDW contributions.

Therefore, the total energy, momentum, angular momentum, and boost momentum of light are given by
\begin{align}
 E_\mathrm{MP} &=\int W_\mathrm{MP}d^3r,\hspace{0.5cm} \mathbf{p}_\mathrm{MP} =\int \mathbf{G}_\mathrm{MP}d^3r,\\
 \mathbf{J}_\mathrm{MP} &=\int \boldsymbol{\mathcal{J}}_\mathrm{MP}d^3r,\hspace{0.4cm} \mathbf{N}_\mathrm{MP} =\int \boldsymbol{\mathcal{N}}_\mathrm{MP}d^3r.
\end{align}

The Lorentz transformation of the MP energy-momentum four-vector is given by
\begin{align}
 E_\mathrm{MP}' &=\gamma(E_\mathrm{MP}-\mathbf{v}\cdot\mathbf{p}_\mathrm{MP}),\\
 \mathbf{p}_\mathrm{MP}' &=\mathbf{p}_{\mathrm{MP},\perp}+\gamma(\mathbf{p}_{\mathrm{MP},\parallel}-\frac{1}{c^2}E_\mathrm{MP}\mathbf{v}).
\end{align}
The Lorentz transformation of $\mathbf{J}_\mathrm{MP}$ and $\mathbf{N}_\mathrm{MP}$, given by
\begin{align}
 \mathbf{J}_\mathrm{MP}' &=\mathbf{J}_{\mathrm{MP},\parallel}+\gamma(\mathbf{J}_{\mathrm{MP},\perp}+\mathbf{v}\times\mathbf{N}_\mathrm{MP}),\\
 \mathbf{N}_\mathrm{MP}' &=\mathbf{N}_{\mathrm{MP},\parallel}+\gamma(\mathbf{N}_{\mathrm{MP},\perp}-\frac{1}{c^2}\mathbf{v}\times\mathbf{J}_\mathrm{MP}),
\end{align}
is similar to the transformation of the fields $\mathbf{E}$ and $\mathbf{B}$ in Eqs.~\eqref{eq:LorentzE} and \eqref{eq:LorentzB}.

\section{\label{sec:comparison}Comparison of the MP and Minkowski SEM tensors}

Next, we compare the MP theory of light with the conventional Minkowski SEM tensor formulation, which has previously been claimed to be the correct canonical formulation of the field and material responses in nondispersive media \cite{Kemp2017,Barnett2010b,Barnett2010a,Bliokh2017a,Bliokh2017b,Brevik2018b}. In particular, we concentrate on a few selected points related to the SEM tensor to illustrate weaknesses of the Minkowski SEM theory, and to describe how these weaknesses are not present in the MP theory of light. A summary of the comparison of the MP SEM tensor and the Minkowski SEM tensor can be found in Table \ref{tbl:comparison}.

\subsection{Minkowski SEM tensor}

The SEM tensor in the conventional Minkowski SEM theory is given by \cite{Pfeifer2007}
\begin{align}
 &\mathbf{T}_\mathrm{M}\nonumber\\
 &\!=\mathbf{T}_\mathrm{field}+\bigg[\begin{array}{cc}
  0 & \mathbf{0}\\
  c\mathbf{D}\!\times\!\mathbf{B}-\frac{1}{c}\mathbf{E}\!\times\!\mathbf{H} & \mathbf{0}
  \end{array}\bigg]\nonumber\\
&\!=\!\!\bigg[\begin{array}{cc}
  \!\!\frac{1}{2}(\mathbf{E}\!\cdot\!\mathbf{D}\!+\!\mathbf{H}\!\cdot\!\mathbf{B}) & \frac{1}{c}(\mathbf{E}\!\times\!\mathbf{H})^T\\
  c\mathbf{D}\!\times\!\mathbf{B} & \frac{1}{2}(\mathbf{E}\!\cdot\!\mathbf{D}\!+\!\mathbf{H}\!\cdot\!\mathbf{B})\mathbf{I}\!-\!\mathbf{E}\otimes\mathbf{D}\!-\!\mathbf{H}\otimes\mathbf{B}\!
  \end{array}\bigg].
\label{eq:tensorMinkowski}
\end{align}
We can see that the difference of the Minkowski momentum density $\mathbf{G}_\mathrm{M}=\mathbf{D}\times\mathbf{B}$ and the Abraham momentum density $\mathbf{G}_\mathrm{A}=\mathbf{E}\times\mathbf{H}/c^2$ has been added to the SEM tensor of the field in an asymmetric way in the left column. In the L frame, this difference is equal to the momentum density of the MDW \cite{Partanen2017c}. The importance of including the momentum of the medium in the theory has been understood widely in the literature and the Minkowski SEM tensor has been presented as a tool to account for the medium part of the momentum \cite{Brevik1979}.

Also, note that in the L frame, the Minkowski SEM tensor can be formed from the MP SEM tensor in Eq.~\eqref{eq:tensormp} by using the substitutions $\mathbf{G}_\mathrm{MP}\rightarrow \mathbf{G}_\mathrm{MP}-\rho_\mathrm{MDW}\mathbf{v}_\mathrm{l}=\mathbf{G}_\mathrm{field}$ and $W_\mathrm{MP}\rightarrow W_\mathrm{MP}-\rho_\mathrm{MDW}c^2=W_\mathrm{field}$ in the first row. To obtain equal expressions, one must also use Eqs.~\eqref{eq:mdwrestframemomentum} and \eqref{eq:mdwrestframestress}, which present the MDW momentum density and the MDW stress tensor in terms of the field quantities in the L frame.

\subsection{Fulfillment of the conservation laws}

The Minkowski SEM tensor can be shown to satisfy the Lorentz transformation in Eq.~\eqref{eq:LorentzT} when the fields transform according to the relations in Eqs.~\eqref{eq:LorentzE}--\eqref{eq:LorentzB} \cite{Kemp2017}. One can also note that the Minkowski SEM tensor in Eq.~\eqref{eq:tensorMinkowski} satisfies the conservation laws of energy and momentum in Eqs.~\eqref{eq:conservationphi} and \eqref{eq:conservationf} if the four-divergences are taken from its row vectors. However, due to the asymmetry, the Minkowski SEM tensor cannot be used to write a consistent AMD tensor through its definition in Eq.\eqref{eq:angularmomentumdensity}. Consequently, the conservation law of angular momentum in Eq.~\eqref{eq:angularmomentumconservation} is not satisfied.

In contrast, the MP formulation of light is expected to be the correct covariant formulation of electrodynamics since, in addition to the Lorentz covariance, it fulfills the conservation laws in Eq.~\eqref{eq:conservation} in the full form including also the diagonal symmetry related to the conservation law of angular momentum in Eq.~\eqref{eq:angularmomentumconservation}.

\begin{figure*}
\centering
\includegraphics[width=\textwidth]{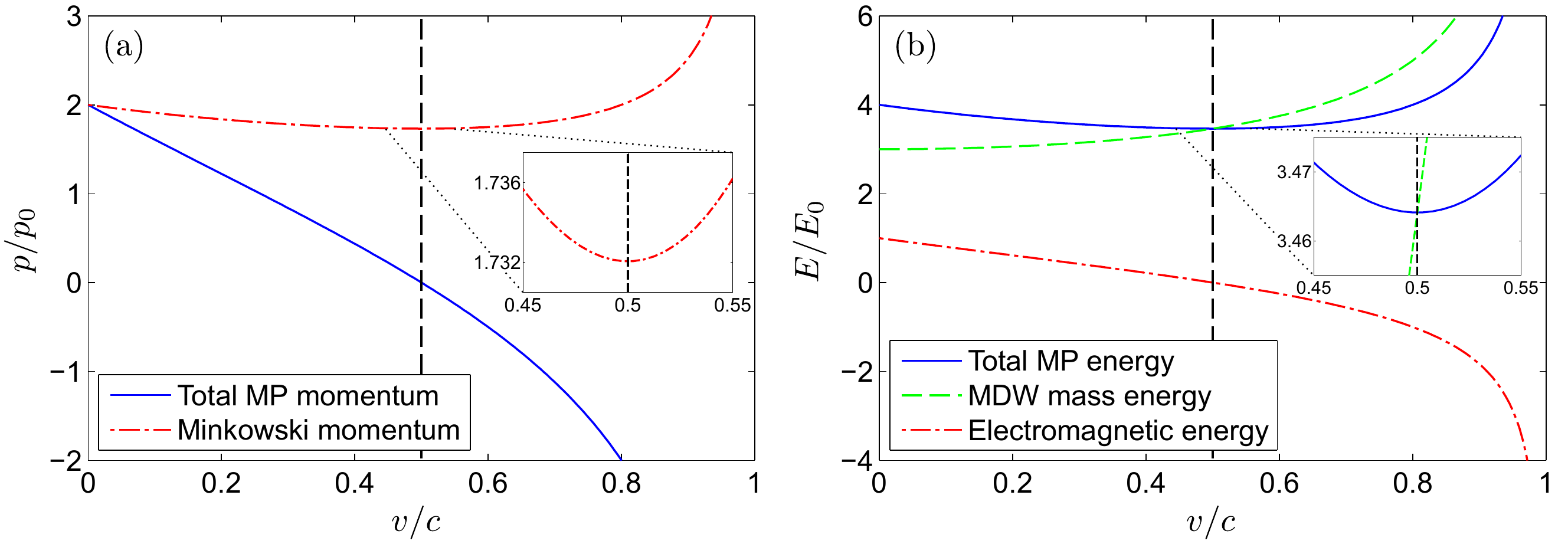}
\vspace{-0.5cm}
\caption{\label{fig:momentaandenergies}
(a) The momentum of the coupled MP state of the field and the atomic MDW and the momentum of light in the Minkowski SEM theory as a function of the relative velocity of the observer parallel to the propagation velocity of light. (b) The electromagnetic energy, MDW mass energy, and the total MP energy as a function of the relative velocity of the observer parallel to the propagation velocity of light. The momenta and energies have been normalized by $p_0$ and $E_0$, the momentum and energy of a light pulse in vacuum in the L frame, and the observer velocity is relative to the L frame. The refractive index in this example is $n=2$ in the L frame where $v/c=0$. The vertical dashed lines correspond to the velocity of light in the medium.}
\end{figure*}

\subsection{Expression of the momentum of light}

Comparing the right-hand sides of the first two lines of Eq.~\eqref{eq:Abrahamforce2}, we find the following expression for the total MP momentum density, given by
\begin{equation}
 \mathbf{G}_\mathrm{MP}=\frac{\mathbf{E}\times\mathbf{H}}{c^2}+\rho_\mathrm{MDW}\mathbf{v}_\mathrm{l}=\mathbf{D}\times\mathbf{B}+\rho_\mathrm{MDW}\mathbf{v}_\mathrm{a}.
\end{equation}
From this equation, it is evident that the MP momentum density $\mathbf{G}_\mathrm{MP}=\mathbf{E}\times\mathbf{H}/c^2+\rho_\mathrm{MDW}\mathbf{v}_\mathrm{l}$ is equal to the Minkowski momentum density $\mathbf{G}_\mathrm{M}=\mathbf{D}\times\mathbf{B}$ only in the L frame, where we can set $\rho_\mathrm{MDW}\mathbf{v}_\mathrm{a}\approx 0$ due to the second-order total dependence of this term on the small atomic velocity $\mathbf{v}_\mathrm{a}$ in the L frame.

In moving reference frames, one has $\mathbf{G}_\mathrm{M}\neq\mathbf{G}_\mathrm{MP}$ as illustrated in Fig.~\ref{fig:momentaandenergies}(a), where the MP and Minkowski momenta, i.e., volume integrals of the momentum densities, are presented as a function of the relative velocity between the observer and the L frame when the observer is moving parallel to light. In particular, the Minkowski momentum obtains a nonzero minimum value in the frame that propagates with the velocity of light in the medium. Thus, the Minkowski momentum is always pointing to the positive direction independently of the velocity of the observer. This applies even in the case in which the observer velocity exceeds the velocity of light in the medium so that light is propagating backward in the inertial frame of the observer. Therefore, this result seems to be against all known measurements of the total momentum of any systems of particles and fields. One should ask why light in a medium would behave in this kind of an odd way.

In contrast, the MP momentum in Fig.~\ref{fig:momentaandenergies}(a) is seen to become zero in the frame that propagates with the velocity of light in the medium (R frame) and, at larger velocities, it points backward just as expected for any particle or quasiparticle with a positive rest mass. Essentially, this is the natural result supported by all previous direct momentum measurements of any systems of particles and fields.

The non-equivalence of the MP and Minkowski momenta strongly suggests that the Minkowski momentum does not have any universal physical meaning as the total momentum of light. In particular, in contrast to previous discussions \cite{Kemp2017,Kemp2015,Barnett2010b,Barnett2010a,Bliokh2017a,Bliokh2017b}, the Minkowski momentum is not the correct momentum of light in moving media as it deviates from the total momentum of the field and the MDW, which is the conserved momentum corresponding to the full relativistically consistent SEM tensor of the MP in Eq.~\eqref{eq:tensormp}.

\subsection{Doppler shift and the rest frame of light}

Using the Lorentz transformation in Eqs.~\eqref{eq:LorentzE}--\eqref{eq:LorentzB} for the field energy in Eq.~\eqref{eq:energyfield}, one obtains the conventional Doppler shift of the electromagnetic energy. As a special case, the electromagnetic energy of light becomes zero in the R frame, which propagates with the velocity of light in the medium. Both the conventional Minkowski SEM theory and the MP theory of light lead to this result as they use the same conventional expression for the electromagnetic energy density given in Eq.~\eqref{eq:energyfield}. The Doppler-shifted electromagnetic energy is presented in Fig.~\ref{fig:momentaandenergies}(b) as a function of the relative velocity between the observer and the L frame when the observer is moving parallel to light.

By comparing Figs.~\ref{fig:momentaandenergies}(a) and \ref{fig:momentaandenergies}(b), we observe yet another fundamental unphysical property of the Minkowski SEM tensor. For the Minkowski SEM tensor, the coexistence of the nonzero momentum in Fig.~\ref{fig:momentaandenergies}(a) with the zero electromagnetic energy in Fig.~\ref{fig:momentaandenergies}(b) in the R frame, raises a natural question what carries this nonzero momentum. That the Minkowski SEM tensor predicts the existence of an object, which has zero energy but nonzero momentum, is in striking contradiction with our present understanding of physics.

In contrast, in the MP theory of light, the total momentum of the MP is zero and the total energy of the MP obtains its minimum value in the R frame, which is also the rest frame of the coupled system of the field and the MDW. This is in full agreement with the STR where a coupled system has zero momentum and minimum energy in its rest frame.

In Fig.~\ref{fig:momentaandenergies}(b), one can also see an interesting effect that the electromagnetic energy becomes negative when the observer is moving faster than the velocity of light in the medium. This effect that is also present in the classical Doppler shift of sound means that wave fronts are moving in opposite directions in the frame of the observer. In the case of light, this is only possible as a part of the coupled MP state whose total energy is positive and larger than the atomic rest mass energy of the MDW. However, for the conventional Minkowski SEM theory, this is problematic since there is no positive energy contribution of the MDW present, and thus the total energy of light is negative, which is unphysical. Effective negative energy would be possible for a quasiparticle that corresponds to a hole in the surrounding energy density, but this is not the case with the conventional Minkowski SEM theory, where the mass density of the medium is not disturbed.

\subsection{Relativistic energy-momentum relation and the rest mass}

According to the STR, the relativistic energy-momentum relation of a particle or any system of particles and fields with total energy $E$, momentum $p$, and rest mass $m_0$ reads $E^2-p^2c^2=(m_0c^2)^2$. This equation is fundamentally related to the four-vector property of the energy and momentum and applies without exceptions to particles and fields with or without a rest mass. In the Minkowski SEM theory, a classical light pulse has, in the L frame, momentum $p_\mathrm{M}=nE_\mathrm{field}/c$ corresponding to the electromagnetic energy $E_\mathrm{field}$. This corresponds to an imaginary rest mass $m_0=i\sqrt{n^2-1}\,E_\mathrm{field}/c^2$. Particles or fields with imaginary rest mass are not known to exist in nature. Thus, the Minkowski SEM theory is in contradiction with the fundamental principles of the STR.

In contrast, in the MP theory of light, the total energy of the MP is the sum of the electromagnetic energy $E_\mathrm{field}$ and the MDW mass energy $\delta Mc^2$.
In the L frame, $\delta Mc^2=(n^2-1)E_\mathrm{field}$ and the total momentum of the field and the MDW is correspondingly $p_\mathrm{MP}=nE_\mathrm{field}/c$. Thus, we obtain a positive rest mass $m_0=n\sqrt{n^2-1}\,E_\mathrm{field}/c^2$ for the coupled MP state of light in a medium. This rest mass is obtained by the Lorentz transformation from the MDW mass $\delta M$ and it is consistent with the classical OCD simulations of the propagation of light in a medium as detailed in Ref.~\cite{Partanen2017c}. Figure \ref{fig:momentaandenergies}(b) shows that the minimum value of the MP energy is obtained in the R frame. This minimum value corresponds to the MP rest energy $E_\mathrm{MP,0}=m_0c^2$.

\subsection{Constant center of energy velocity law of an isolated system}

As detailed in Ref.~\cite{Partanen2017c}, the mass transfer of the MDW is necessary for the fulfillment of the constant center of energy velocity (CEV) law of an isolated system, which is commonly known as Newton's first law. This law is violated in the conventional Minkowski SEM theory having no atomic mass transfer.

\subsection{Comment on the use of a curved metric in some previous works}

In some previous works \cite{Gordon1923,Leonhardt2006b,Leonhardt2010,Obukhov2008}, the problems of the conventional Minkowski SEM theory have been artificially solved by introducing the Gordon metric, which depends on the permittivity and permeability of materials. Using the equivalence principle of the general theory of relativity, this metric corresponds to artificial gravitational fields that are not physically true in the sense of the general theory of relativity. Therefore, these works do not solve the problem of formulating the covariant theory of electrodynamics in the space-time whose metric is only modified by true gravitational fields. In contrast, the MP theory of light gives the covariant behavior of the SEM tensor, the correct symmetry properties, the correct conservation laws, and the dynamical equations of the field and the matter without artificial curved metrics.

\subsection{Comment on accounting for the SEM tensor of the medium in previous works}

In many previous works, it has been concluded that the SEM tensor of the medium must be used together with the SEM tensor of the electromagnetic field to describe the propagation of light in a medium \cite{Pfeifer2007,Mikura1976,Ramos2015}. These works typically lead to a complicated form for the total SEM tensor of the field and matter; see, e.g., Eq.~(34) of Ref.~\cite{Pfeifer2007}. The conventional Minkowski SEM tensor and its material counterpart are obtained only in the nonrelativistic limit in Refs.~\cite{Pfeifer2007,Mikura1976}; see, e.g., Eqs.~(42)--(43) of Ref.~\cite{Pfeifer2007}. Thus, the division of the total SEM tensor into the field and the medium parts in these works does not seem to be both unique and form-invariant between different inertial frames. This reported separation is even argued to be arbitrary in Ref.~\cite{Pfeifer2007}.

\begin{table*}
 \setlength{\tabcolsep}{5.0pt}
 \renewcommand{\arraystretch}{2.4}
 \caption{\label{tbl:comparison}
 Comparison of the Minkowski SEM tensor and the MP SEM tensor formalisms.}
\begin{tabular}{>{\raggedright\arraybackslash}p{3.2cm}>{\raggedright\arraybackslash}p{6.8cm}>{\raggedright\arraybackslash}p{6.8cm}}
   \hline\hline
   Required physical property, symmetry, or invariance & Minkowski SEM tensor$^\mathrm{(a}$ & Mass-polariton SEM tensor$^\mathrm{(b}$ \\[4pt]
   \hline\hline
Form-invariance between inertial frames & Fulfilled for the SEM tensor. Not applicable for the related AM tensor, which violates the conservation law of angular momentum. & Fulfilled for the SEM tensor and for the related AM tensor. \\
\hline
Lorentz transformation of the tensor components & Fulfilled for the SEM tensor. Not applicable for the related AM tensor. & Fulfilled for the SEM tensor and for the related AM tensor. Not fulfilled for the field and the MDW parts that are not isolated due to their coupling through the Abraham force. \\
\hline
   Conservation of angular momentum & The Minkowski SEM tensor cannot be used to write an AMD tensor whose four-divergence is zero. Thus, angular momentum is not conserved. & One can write an AMD tensor whose four-divergence is zero. Thus, angular momentum is conserved.\\
\hline
Emergence of the dynamical equations from the SEM tensor & Field dynamics obtained from the SEM tensor, but no dynamics at all for the medium. & Fully consistent dynamics both for the field and the medium. \\
\hline
The law of action and counteraction (Newton's third law) between the field and the medium & The law of action and counteraction not applicable since there is no force-based coupling between the field and the medium. & The law of action and counteraction fulfilled: $\partial_\beta(T_\mathrm{MDW})^{\alpha\beta}=-\partial_\beta(T_\mathrm{field})^{\alpha\beta}$.\\
\hline
Need of an artificial curved metric to make the SEM tensor symmetric & One must introduce an artificial curved metric \cite{Gordon1923}. Due to the equivalence principle of the general relativity, this is equivalent to introducing artificial gravitational fields. & No artificial curved metric and equivalent artificial gravitational fields needed. \\
\hline
   Total energy, momentum, and rest mass in the R frame & $E_\mathrm{M}^{(\mathrm{R})}=0$, $p_\mathrm{M}^{(\mathrm{R})}=\sqrt{n^2-1}\,E_\mathrm{field}^{(\mathrm{L})}/c$, $m_0=i\sqrt{n^2-1}\,E_\mathrm{field}^{(\mathrm{L})}/c^2$. An object with zero energy, nonzero momentum, and imaginary rest mass is against our fundamental understanding of physics and is not to be found in nature. & $E_\mathrm{MP}^{(\mathrm{R})}=n\sqrt{n^2-1}\,E_\mathrm{field}^{(\mathrm{L})}$, $p_\mathrm{MP}^{(\mathrm{R})}=0$, $m_0=n\sqrt{n^2-1}\,E_\mathrm{field}^{(\mathrm{L})}/c^2$. In accordance with the STR, the minimum of the total energy of a light pulse is obtained in the R frame, where the field energy and the total momentum are zero. The origin of the positive rest mass well understood.\\
\hline
   Constant CEV of a light pulse (Newton's first law) at material interfaces & Accounting for the recoil force, the momentum is conserved. The constant CEV law violated at material interfaces. & Accounting for the recoil force, the momentum is conserved. The constant CEV law fulfilled at material interfaces.\\
   \hline\hline
 \end{tabular}
\flushleft
a) Some works \cite{Pfeifer2007} add a material counterpart to the Minkowski SEM tensor, which is neglected here.\\
b) In this work, we neglect the losses related to the strain energies that are left in the medium due to the displacement of atoms by the optical force \cite{Partanen2017c}. Therefore, the MP SEM tensor, as defined in the present work, does not describe the elastic relaxation of the medium after a light pulse.
\end{table*}

\pagebreak
\clearpage

In contrast, in the MP theory, the division of the energy and momentum between the field and the atomic MDW is accurately described in a unique, form-invariant, and physically transparent way in any inertial frame. In the classical MP theory of light discussed in this work, the separate field and the MDW parts of the coupled MP state of light are unambiguously defined and independently experimentally measurable. Thus, there cannot be any arbitrariness in the sharing of energy, momentum, or angular momentum between these classical objects.

\section{\label{sec:conclusions}Conclusions}

In conclusion, we have proved the Lorentz covariance of the MP theory of light. In contrast to the conventional Minkowski SEM theory, the MP theory accounts for the field-driven atomic MDW mass, momentum, and stress terms. Consequently, the MP SEM tensor is diagonally symmetric in contrast to the conventional Minkowski SEM tensor. We have also discussed how accounting for the MDW terms solves several weaknesses that are present in the conventional Minkowski SEM theory. Remarkably, in contrast to previous suggestions \cite{Kemp2017,Kemp2015,Barnett2010b,Barnett2010a,Bliokh2017a,Bliokh2017b}, our results strongly suggest that the Minkowski momentum is not the universally correct momentum of light as it is found to be equal to the total momentum of the field and the MDW only in the L frame. This result also has far-reaching consequences in the theory of optical angular momentum, where the atomic MDW plays a substantial role \cite{Partanen2018a}.

\begin{acknowledgments}
This work has been funded in part by the Academy of Finland under Contracts No.~287074 and No.~318197. We want to thank K.~Y.~Bliokh for notifying us on the terminology related to the boost momentum term that is used in previous optics literature. We also want to thank T.~Po\ifmmode\check{z}\else\v{z}\fi{}ar and N.~G.~C.~Astrath for discussions on different SEM tensor formalisms, and H.~Lee, H.~Choi, and K.~Oh for discussions on optical forces and the momentum of light.
\end{acknowledgments}

\end{document}